\documentclass[envcountsect,envcountsame,runningheads,a4paper]{llncs}

\usepackage{main}
\usepackage[subtle,tracking=normal]{savetrees}

\usepackage{orcidlink}
\hypersetup{pdfborder={0 0 0}}

\usepackage{graphbox} 

\AtBeginEnvironment{alltt}{\setlength{\topsep}{2pt}}

\begin{document}

\title{Formal Verification of \\ Physical Layer Security Protocols for \\Next-Generation Communication Networks}
\titlerunning{Formal verification of PLS protocols}

\author{Kangfeng Ye \inst{1}\orcidID{0000-0003-2460-7926} \and 
Roberto Metere \inst{1}\orcidID{0000-0001-6992-4285}  \and
Jim Woodcock \inst{2,3,1}\orcidID{0000-0001-7955-2702}
\and 
Poonam Yadav \inst{1}\orcidID{0000-0003-0169-0704}
}

\institute{%
University of York, UK \\\email{\{kangfeng.ye,roberto.metere,jim.woodcock,poonam.yadav\}@york.ac.uk} 
\and
Southwest University, China
\and
Aarhus University, Denmark
}

\authorrunning{Ye et al.}

\maketitle

\begin{abstract}
    Formal verification is crucial for ensuring the robustness of security protocols against adversarial attacks.
    The Needham-Schroeder protocol, a foundational authentication mechanism, has been extensively studied, including its integration with Physical Layer Security (PLS) techniques such as watermarking and jamming.
    Recent research has used ProVerif to verify these mechanisms in terms of secrecy. However, the ProVerif-based approach limits the ability to improve understanding of security beyond verification results. To overcome these limitations, we re-model the same protocol using an Isabelle formalism that generates sound animation, enabling interactive and automated formal verification of security protocols.
    Our modelling and verification framework is generic and highly configurable, supporting both cryptography and PLS. For the same protocol, we have conducted a comprehensive analysis (secrecy and authenticity in four different eavesdropper locations under both passive and active attacks) using our new web interface. 
    Our findings not only successfully reproduce and reinforce previous results on secrecy but also reveal an uncommon but expected outcome: authenticity is preserved across all examined scenarios, even in cases where secrecy is compromised. 
    We have proposed a PLS-based Diffie-Hellman protocol that integrates watermarking and jamming, and our analysis shows that it is secure for deriving a session key with required authentication. 
    These highlight the advantages of our novel approach, demonstrating its robustness in formally verifying security properties beyond conventional methods.

\end{abstract}

\section{Introduction}
\label{sec:intro}
Classic cryptographic methods~\cite{Menezes1996} assume the computational hardness of inverting one-way trapdoor functions (e.g., prime factoring or discrete logarithm) and the limited computational power of adversaries. 
Adversaries that can use a sufficiently powerful quantum computer would employ techniques based on Shor's algorithm~\cite{Shor1994} to break (invert) the above functions efficiently.
As quantum computers are currently limited in the number of qubits required to attack deployed systems, a temporary mitigation is to use longer keys, which necessitate more complex key generation and management. 
This extra complexity 
will also need more computational power and cause delays, translating into a waste of resources and making it difficult or even impossible to be deployed in resource-constrained, power-limited, or latency-sensitive devices, 
typical of next-generation networks, such as 6G. 
Modern networks, indeed, demand enhanced security requirements~\cite{Nguyen2021} for highly diverse applications.
An example are demands for strict quality of service (QoS), such as ultra-low latency and extreme high reliability (e.g., for extended reality and remote surgery), highly dynamic and heterogeneous wireless networks, massive number of low-cost Internet of Things (IoT) sensor devices, and evolving threats from Machine Learning (ML) and Artificial Intelligent (AI) in 6G, Physical Layer Security (PLS)~\cite{Mitev2023,Zhang2024} is emerging as a complementary approach to add an extra layer security.
In this context, providing security guarantees at the physical layer significantly reduces the need for the above complications.

PLS aims to secure wireless communication at the physical layer (the lowest network layer) by exploiting the inherent characteristics and randomness of the wireless communication channel itself, such as fading, noise, interference, dispersion, and diversity. 
PLS 
is based on information-theoretic security or unconditional security, a paradigm of security that focuses on the fundamental limits of information leakage as defined by information theory~\cite{Shannon1948}.
It aims to provide security that is provably unbreakable regardless of the eavesdropper's computational resources, such as Shannon's perfect secrecy~\cite{Shannon1948} 
, weak secrecy~\cite{Wyner1975} 
and strong secrecy~\cite{Csiszar1996} 
(or statistical independence and asymptotically secure).
PLS can resist some physical layer attacks such as spoofing, eavesdropping, and jamming (while conventional cryptography is not able to do so), can be key-less~\cite{Kumar2022} (so no complex key management), and is quantum-resistant (because it is not based on computational hardness).
Additionally, it has low computational complexity and latency, making it intrinsically suitable for IoT devices and heterogeneous wireless networks in 6G. 

While a large amount of work in PLS focuses on 
secrecy, PLS techniques for achieving authentication have also received considerable attention. For example, Soderi et al.\cite{Soderi2017} proposed the watermark-based blind physical layer security (WBPLSec) approach to achieve both secrecy and message integrity (i.e., authenticity with proper protocols). The method combines watermarking\cite{Cox1999,Li2013}, where the sender embeds a watermark or authentication information in a content signal to create a covert channel based on a shared secret, with jamming, where the legitimate receiver selectively interferes with parts of the signal so that only it can recover the message while eavesdroppers cannot. 

Notwithstanding its considerable potential, formal verification of PLS mechanisms constitutes an underdeveloped area of research.
Traditional cryptographic security protocols have benefited from formal verification with useful tools such as FDR~\cite{GibsonRobinson2014}, Isabelle~\cite{Nipkow2002}, Maude-NPA~\cite{Escobar2006}, AVISPA~\cite{Armando2005}, ProVerif~\cite{Blanchet2016}, and Tamarin-prover~\cite{Basin2017}, 
\begin{enumerate*}[label=(\arabic*)]
    \item to discover attacks~\cite{Lowe1996,Arapinis2012};
    \item to identify missing or weak assumptions~\cite{Basin2018};
    \item to propose fixes or improvements to protocols~\cite{Lowe1996,Basin2018,Miller2022}; and 
    \item to guarantee correctness~\cite{Bhargavan2017,Basin2022}.
\end{enumerate*}
These formal security analysis methods primarily focus on higher-layer cryptographic protocols. PLS protocols, however, operate in the lower-layer physical layers and use different security mechanisms from cryptography. Therefore, it is highly beneficial to formally verify PLS protocols before their vulnerabilities and missing assumptions are discovered after their deployment.

A recent work~\cite{Costa2023} has 
leveraged ProVerif~\cite{Blanchet2016}, a security verification tool, to verify the secrecy of a WBPLSec-based variant (called NSWJ) of the Needham-Schroeder protocol~\cite{Needham1978}.
Their verification approach, however, is carried out by formal verification experts in that specific language.
\emph{Protocol designers still cannot use the approach to inspect and verify the protocols by themselves.}
In this regard, our recent work~\cite{Ye2024a} provides more accessibility to formal verification through sound animation.
This limits the intervention of experts and allows designers to either manually explore or automatically verify cryptographic protocols.
In such a workflow, formal experts use a variant of Communicating Sequential Processes (CSP)~\cite{Hoare1985,Roscoe2011}, called the Interaction Trees~\cite{Xia2019} (ITrees) based CSP (ITrCSP)~\cite{Foster2021,Ye2024}, to model security protocols in the theorem prover Isabelle/HOL~\cite{Nipkow2002}.
Then, Isabelle can automatically generate Haskell code that, when compiled, provides a lightweight model checker or animator requiring no more expert intervention.
However, this work only supports the Dolev-Yao attack model~\cite{Dolev1983}, the ad-hoc framework for each protocol (each protocol has a different fundamental message theory), and a terminal interface (users at least need a PC or workstation to run animators).
Physical layer security is defined in terms of a different attacker model, so we extended \cite{Ye2024a} to model it.

Based on the context, this work is guided by the following research questions: 
\begin{enumerate*}[label={\textbf{RQ\arabic*}}]
    \item(PLS): How to extend the ITrCSP-based framework to support PLS and its related attack model for verification? \label{RQ1:pls}
    \item(Generic and generalisation) Can the ITrCSP-based framework be general to model all different attack models and security protocols in a generic message theory?\label{RQ2:general}
    \item(Case studies) What different insights can we get if we conduct a more thorough analysis of NSWJ? Can WBPLSec be used to implement the Diffie-Hellman algorithm? What benefits can PLS bring? \label{RQ3:cases}
    \item(Accessibility) Can the framework be made even more accessible to a wider range of stakeholders, such as non-professional groups like students and the general public? \label{RQ4:accessibility}
\end{enumerate*}

In this paper, we present our work that addresses these questions. Our novel contributions here are as follows: 
\begin{enumerate*}[label=(\arabic*)]
    \item a general sound animation framework to support both classic cryptography and PLS techniques, built on a generic and polymorphic message theory;
    \item a more scalable implementation of intruder's message inference rules by constraining build-up messages into all the possible messages that legitimate agents can receive, instead of the naive implementation of~\cite{Ye2024a} to build up all possible messages from inference rules; 
    \item a comprehensive analysis of NSWJ by considering not only secrecy but also authenticity in both passive and active attacks when the adversary could be within or out of the jamming ranges of each legitimate participant, to confirm the result of secrecy with~\cite{Costa2023} but contradicts the interpretation of authenticity from~\cite{Costa2023};
    \item a new WBPLSec-based DH protocol (DHWJ) to fix the man-in-the-middle attack and achieve authenticity;
    \item a new web interface to make our animation more accessible and user-friendly, in addition to the terminal interface.
\end{enumerate*}
All definitions in this paper are mechanised and show an icon (\isalogo) that links to the corresponding repository artefacts\footnote{We assume basic knowledge of Isabelle/HOL from interested readers to understand these definitions and theorems.}.

The remainder of this paper is organised as follows. %
We review related work in \Cref{sec:related}, introduce sound animation and WBPLSec in \Cref{sec:background}, and present the modelling of security protocols in our generic framework in~\Cref{sec:modelling}. And in~\Cref{sec:nsdh}, we describe the modelling of NSWJ and DHWJ in our framework, then present the web interface and discuss verification results in~\Cref{sec:verification}. %
Finally, in~\Cref{sec:concl}, we conclude and discuss future work.

\section{Related work}
\label{sec:related}

\textbf{Formal verification of security protocols.} 
Model checking and theorem proving tools, such as Casper~\cite{Lowe1997a}, FDR~\cite{Lowe1996,GibsonRobinson2014}, Isabelle~\cite{Nipkow2002,Paulson1997}, Maude-NPA~\cite{Escobar2006}, AVISPA~\cite{Armando2005}, ProVerif~\cite{Blanchet2016}, and Tamarin-prover~\cite{Basin2017}, have been commonly used to formally verify security protocols. These tools require users to have an in-depth level of knowledge in formal verification, and thus are not accessible to general protocol designers. Instead, the work presented in this paper provides a more accessible approach using sound animation, allowing protocol designers to interact, inspect, and verify protocols through interfaces, as well as to perform automated verification. 
\changed[\Cn{2}{5}]{Similar to our work, SPAN~\cite{Boichut2007} is an animation tool and can be used by designers. But its soundness is not guaranteed.} 

\textbf{Formal verification of PLS.} Formal verification for physical-layer security protocols is a relatively new area. To the best of our knowledge, the work from Costa et al.~\cite{Costa2023} is the only one that focuses on this topic. In their work, the standard ProVerif language was extended with functions to model and analyse WBPLSec-based protocols symbolically. Furthermore, an attack model for WBPLSec was implemented, and their work demonstrated that NSWJ with WBPLSec is secure (in terms of secrecy) against active attackers within the jamming range and insecure otherwise. 
Their work formally demonstrates the sufficiency of watermarking and jamming to ensure secrecy and authenticity.
Inspired by~\cite{Costa2023}, our work verifies WBPLSec and its based NSWJ protocol, in the same scenarios and under the same common principles, such as its attack model and assumptions.
Conversely, our work goes beyond to comprehensively analyse all \emph{four scenarios}, where the eavesdropper is located at different positions in terms of Alice's and Bob's jamming ranges, under both \emph{passive and active attacks}. The analysis provides previously undiscovered results: \emph{authenticity is preserved even when the eavesdropper is outside the jamming range (and thus, the loss of secrecy).}
We also propose a new variant (DHWJ) of DH, based on WBPLSec, and conduct a comprehensive analysis of it.
The other benefit of our work, if compared to theirs, is the \emph{accessibility} of verification to protocol designers.
It is worth noting that \emph{the capability to understand and debug protocols} is crucial for protocol designers too, especially in new areas such as PLS.
As a result, we demonstrate how our analysis facilitates a deeper understanding of the protocol, as it contradicts one of the conclusions in~\cite{Costa2023}, as mentioned previously.

\textbf{Sound animation.} For another comparison, ITrCSP has been applied to animate robotic control software~\cite{Ye2024} modelled in RoboChart~\cite{Miyazawa2019,Ye2022} and security protocols~\cite{Ye2024a}, and demonstrate functional correctness through manual interaction and automatic checking. Our work is also based on ITrCSP and extends~\cite{Ye2024a} in three aspects: a general modelling framework, the WBPLSec security protocol, and a web user interface. The sound animation approach presented in~\cite{Ye2024a} primarily focuses on two case studies. Its modelling framework in ITrCSP and Isabelle is not generic. Thus, every protocol has an ad-hoc message theory in Isabelle. Our work presents a generic and polymorphic message theory, enabling all protocols to share a common message framework. Additionally, each protocol can be instantiated for specific configurations. This will be discussed in \Cref{sec:modelling}.
We also extend the modelling framework to support a new attack model for WBPLSec in addition to the Dolev-Yao model. WBPLSec-based protocols, such as NSWJ and DHWJ, can be modelled and verified in the same framework. 
From an interface aspect, we develop a new web interface to allow users to easily configure properties, verify, and view their interactions with models and counterexamples. 


\changed[\Cn{2}{5}]{
\textbf{Summary.} \Cref{table:related_work} summarises the comparison of this work with others in modelling primitives and attack models, soundness, verification, and interfaces. 
    It is worth noting that 
    \begin{enumerate*}[label=(\arabic*)]
        \item the claimed soundness in Taramin and ProVerif is not formally verified, unlike our work where the soundness is guaranteed by Isabelle/HOL; and
        \item Taramin has a web interface for its interactive proof mode, but not intended for designers.
    \end{enumerate*}
\begin{table*}[bt]
    \caption{\changed[\Cn{2}{5}]{Comparison of related work where $\emptycirc[0.5ex]$ for no, $\fullcirc[0.5ex]$ for yes, and $^\ast$ for comments.}}
    \label{table:related_work}\centering
\bgroup
\begin{tabular}{@{}l l cc l cc l c l cccc l cc l@{}}
        \toprule
        \multirow{2}{*}{Related work} & 
        \tablecolsep & 
        \multicolumn{2}{c}{Primitives} & 
        \tablecolsep & 
        \multicolumn{2}{c}{Attack} & 
        \tablecolsep & 
        \multirow{2}{*}{Sound} & 
        \tablecolsep & 
        \multicolumn{4}{c}{Verification} & 
        \tablecolsep &  
        \multicolumn{2}{c}{UI} & 
        \\
        \cline {3-4}
        \cline {6-7}
        \cline {11-14}
        \cline {16-17}
         & &
        Crypt & PLS & &
        DY & PLS & &
        & &
        DBG & UGV & Reach & Feas & &
        Term & Web & 
        \\ 
        \midrule
        Tamarin & &
        \fullcirc[0.5ex] & \emptycirc[0.5ex] & &
        \fullcirc[0.5ex] & \emptycirc[0.5ex] & &
        \fullcirc[0.5ex]$^\ast$ & &
        \emptycirc[0.5ex] & \emptycirc[0.5ex] & \fullcirc[0.5ex] & \emptycirc[0.5ex] & &
        \fullcirc[0.5ex] & \fullcirc[0.5ex]$^\ast$ & 
        \\ 
        ProVerif & &
        \fullcirc[0.5ex] & \emptycirc[0.5ex] & &
        \fullcirc[0.5ex] & \emptycirc[0.5ex] & &
        \fullcirc[0.5ex]$^\ast$ & &
        \emptycirc[0.5ex] & \emptycirc[0.5ex] & \fullcirc[0.5ex] & \emptycirc[0.5ex] & &
        \fullcirc[0.5ex] & \emptycirc[0.5ex] & 
        \\ 
        Costa et al.~\cite{Costa2023} & &
        \fullcirc[0.5ex] & \fullcirc[0.5ex] & &
        \fullcirc[0.5ex] & \fullcirc[0.5ex] & &
        \fullcirc[0.5ex]$^\ast$ & &
        \emptycirc[0.5ex] & \emptycirc[0.5ex] & \fullcirc[0.5ex] & \emptycirc[0.5ex] & &
        \fullcirc[0.5ex] & \emptycirc[0.5ex] & 
        \\ 
        SPAN~\cite{Boichut2007}& &
        \fullcirc[0.5ex] & \emptycirc[0.5ex] & &
        \fullcirc[0.5ex] & \emptycirc[0.5ex] & &
        \emptycirc[0.5ex] & &
        \fullcirc[0.5ex] & \emptycirc[0.5ex] & \fullcirc[0.5ex] & \emptycirc[0.5ex] & &
        \fullcirc[0.5ex] & \fullcirc[0.5ex] & 
        \\ 
        Ye et al.~\cite{Ye2024a}& &
        \fullcirc[0.5ex] & \emptycirc[0.5ex] & &
        \fullcirc[0.5ex] & \emptycirc[0.5ex] & &
        \fullcirc[0.5ex] & &
        \fullcirc[0.5ex] & \fullcirc[0.5ex] & \fullcirc[0.5ex] & \fullcirc[0.5ex] & &
        \fullcirc[0.5ex] & \emptycirc[0.5ex] & 
        \\ 
        This work & &
        \fullcirc[0.5ex] & \fullcirc[0.5ex] & &
        \fullcirc[0.5ex] & \fullcirc[0.5ex] & &
        \fullcirc[0.5ex] & &
        \fullcirc[0.5ex] & \fullcirc[0.5ex] & \fullcirc[0.5ex] & \fullcirc[0.5ex] & &
        \fullcirc[0.5ex] & \fullcirc[0.5ex] & 
        \\ 
        \midrule
        \multicolumn{18}{p{.95\linewidth}}{\textbf{Acronym} 
        Crypt: cryptography;
        Attack: attack models;
        DY: Dolev-Yao; 
        DBG: debugging; 
        Reach: reachability check;
        Feas: feasibility check;
        Term: terminal interface;
        UGV: user-guided verification.
        } \\
        \bottomrule
\end{tabular}
\egroup
\vspace*{-2em}
\end{table*}
}

\section{Background}
\label{sec:background}

\textbf{ITrCSP and Sound animation.}
%
Animation models interactions of reactive systems with their environment through events. We utilise Interaction Trees~\cite{Xia2019} (ITrees), which allow for a formal specification to be associated with both abstract and executable denotational semantics~\cite{Xia2022}, as coinductive tree structures. This enables the construction of denotational and operational semantics, facilitating both reasoning and implementation.
Using ITrees, Foster et al.~\cite{Foster2021,Ye2024} gave semantics to a deterministic version of CSP process algebra~\cite{Hoare1985,Roscoe2011}, called ITree-based CSP or ITrCSP, and mechanised it in Isabelle/HOL~\cite{Nipkow2002} to allow automatic generation of Haskell code from a CSP model for animation. %
The basic processes and operators of ITrCSP are summarised in \cite[Table 1]{Ye2024a}. Further details about their definitions, semantics, and implementation in Isabelle/HOL can be found in \cite{Foster2021,Ye2024}. 

\begin{figure}[t]
    \centering
    \includegraphics[width=0.6\columnwidth]{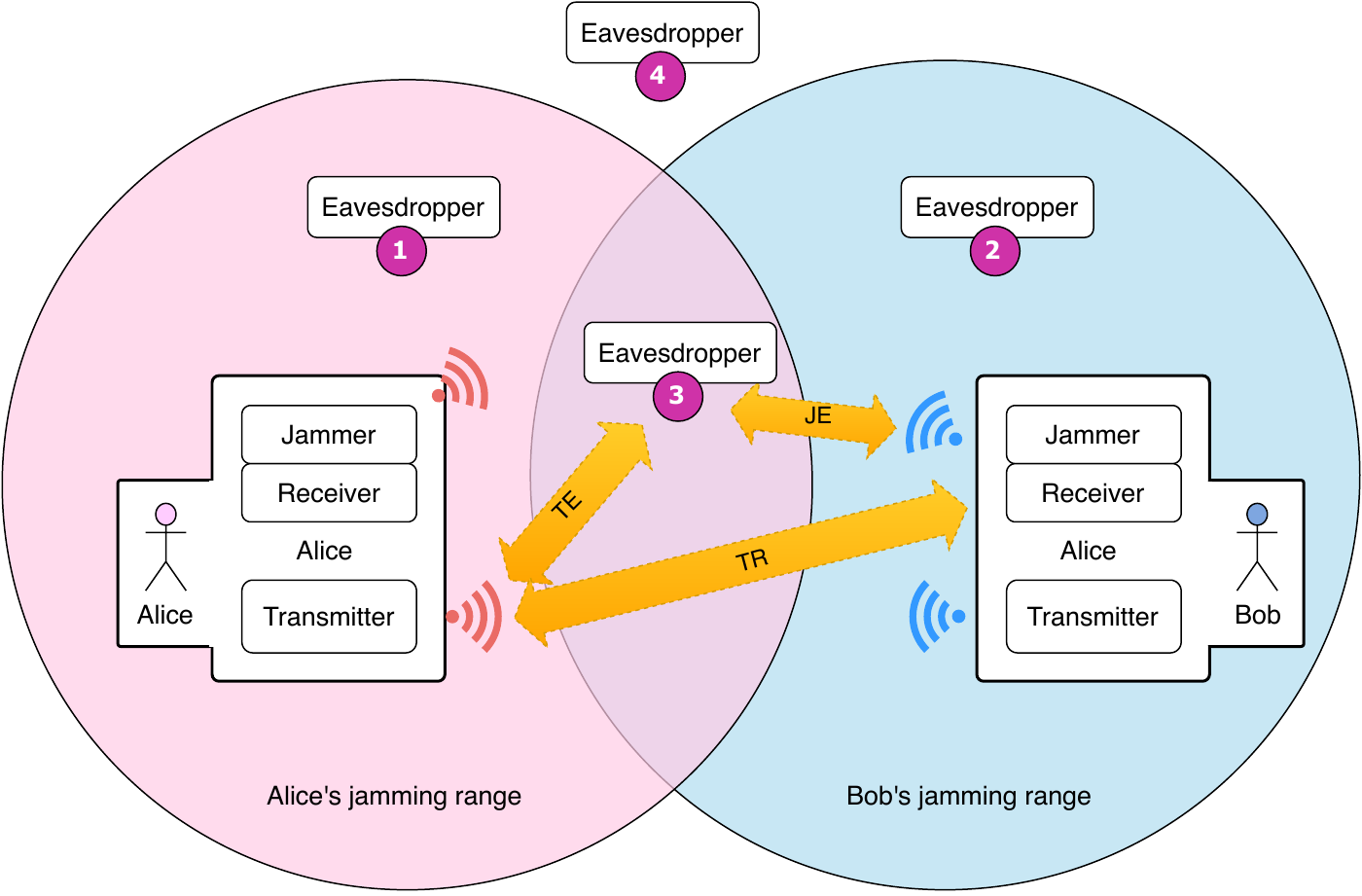}
    \caption{An illustration of WBPLSec with two legitimate agents (Alice and Bob) and an eavesdropper in different spatial locations in terms of the agents' jamming ranges. There are four combinations of locations, denoted as Eve1, Eve2, Eve3, and Eve4. 
    \label{fig:wbplsec}
    }
    \vspace*{-2.0em}
\end{figure}

The animation is sound thanks to ITree's executable semantics and Isabelle's code generator~\cite{Haftmann2010}, which translates executable ITree definitions in HOL logic to target functional languages (such as Haskell). 
{Soundness is not a formally proved property here, but the simple and unbroken link from operational semantics of CSP, to (corecursive) definitions of CSP operators in ITrees, and the final Haskell code through an equational logic-based code generator (all in Isabelle/HOL) ensures its soundness.} We illustrate this link in \Cref{fig:sound_animation} of \Cref{appendix:soundness}.
\textbf{PLS: Watermarking and Jamming. } 
%
Soderi et al.~\cite{Soderi2017} proposed WBPLSec, which combines watermarking~\cite{Cox1999,Li2013} with jamming (based on iJAM~\cite{Gollakota2011} but improved the data rate from half to full) for secure communication. 
During the transmission, the source clear message and the watermarked signal (from a part of the source message) are transmitted in two independent paths, either embedded both signals in a narrow-band channel~\cite{Soderi2017} or two different visible lights (the blue light and the red light) in visible light communication (VLC)~\cite{Soderi2022q}.
\Cref{fig:wbplsec} illustrates the concept of WBPLSec 
schematically.
We consider two legitimate agents, Alice and Bob, and one eavesdropper (Eve) in various locations (Eve 1, 2, 3, and 4) in terms of the jamming ranges (the two large circles in the diagram) of Alice and Bob. Each of the legitimate agents is equipped with a transmitter and a receiver which contains a jammer. There are three kinds of links depicted in the diagram: TR for the transmitter-receiver link, TE for the transmitter-eavesdropper link, and JE for the jammer-eavesdropper link. Considering a scenario where Alice wants to send a confidential message $m$ to Bob, the WBPLSec approach works as follows:
\begin{enumerate}[label=S\arabic*., ref=S\arabic*]
    \item the transmitter of Alice modulates $m$ to get the host regulated signal $x_S$ of length $N$, of which the first $N_w$ samples $x_w$ are selected to use direct sequence spread spectrum (DSSS) for watermarking (by using the pseudo-noise PN code $c_w$ to spread the $x_w$ to decorrelate the host signal with the watermark later) to get the spread-spectrum (SS) watermark $w$; \label{Step:1}
    \item the transmitter embeds $w$ in $x_S$ to get $x'_S$ and sends it;\label{Step:2}
    \item then the receiver Bob uses his jammer to jam $M$ (where $M \leq N_w$) samples over $x'_S$ to get an intentionally interfered and corrupt $\hat{x}_S$;\label{Step:3}
    \item Bob can extract the watermark $\hat{x}_w$ from the received signal (via the TR link) by using an additional DSSS demodulator based on the same PN code $c_w$;\label{Step:4}
    \item Bob's receiver then removes corrupted jammed samples from $\hat{x}_S$ and replaces them with unjammed samples from $\hat{x}_w$;\label{Step:5}
    \item if Eve, such as Eve 2, is within the jamming range of Bob, Eve is not able to construct clean messages from $\hat{x}_S$ (received through the JE link) because she does not know $c_w$;\label{Step:6}
    \item otherwise, if Eve is out of the jamming range of Bob, she can receive the watermarked signal $x'_S$ through the TE link and understand the clean message $x_S$.\label{Step:7}
\end{enumerate}
To ensure secrecy and integrity, Alice and Bob share a common secret, $N$ and $c_w$, beforehand and never send the secret over the network. Bob will use another PN code for his transmitter, which Alice is also aware of in advance.


The security of WBPLSec-based protocols is highly dependent on the location of the eavesdropper. To minimise the information leakage, the location of legitimate agents, their transmission and jamming powers need to be considered. One possible solution is to ensure that jamming fully covers the transmission area. 
This is feasible if the receiver (such as an edge node) is more powerful or visible light in VLC is restricted in an area (for example, by walls). 
Another solution is to use reconfigurable intelligent surfaces (RIS) to dynamically direct transmission signals towards legitimate receivers and then direct jamming signals to cover that transmission area, thereby forming a secure region~\cite{Soderi2022}.

\section{Modelling of Security Protocols}
\label{sec:modelling}
Our framework supports the symbolic modelling of both classic cryptographic and PLS protocols. The formalization of cryptography and the Dolev-Yao attack model~\cite{Dolev1983} have already been described in~\cite{Ye2024a}. For this reason, this section presents our formalization of PLS protocols in excruciating detail, instead of cryptography.


\textbf{Assumptions.} For WBPLSec, we assume
\begin{enumerate}[label={AS\arabic*.}, ref={AS\arabic*}]
    \item As a sender, each legitimate agent has a unique secret code used for watermarking. \label{AS:1}
    \item As a receiver, each legitimate agent knows the unique secret codes which the corresponding senders use for watermarking, and the agent then uses the code to jam the communication from these senders. \label{AS:2}
    \item The intruder (or Eve)\footnote{We use \textit{Intruder} in the modelling of security protocols and only use \textit{Eve} when discussing WBPLSec in general.} does not know or guess the secret codes that legitimate agents use for watermarking and jamming. \label{AS:3}
    \item The intruder may have her code for watermarking, but her code is not known to any legitimate agents. \label{AS:4}
    \item These secret codes are never transmitted over the network. \label{AS:5}
\end{enumerate}

\textbf{Network model.}
A network model describes how different agents and the intruder communicate through proper (secure or public) channels. 
%
For WBPLSec, we employ a different network model than that used for cryptography~\cite{Ye2024a} because watermarking and jamming differ from conventional cryptography: they establish a secure region (based on locations), whereas cryptography enforces security through encryption, hashing, and other primitives.
In the new network model, each legitimate agent sends a watermarked message $wm=\watm{m}{bA_w}$ using its watermarking {bitmask} $bA_w$ (corresponding to the secret code, as with~\cite{Costa2023}).
The intruder has no jammer because she always tends to hear more information instead of interfering, and will receive $wm$.
Its receiver relays $wm$ to Bob, and at the same time, it sends a jammed message $jm=\jam{wm}{bB_j}$ (using Bob's jamming bitmask $bB_j$) to the main body of the intruder if the intruder is within the jamming range of Bob.
Otherwise, the receiver sends $wm$ to the intruder.
This way, the intruder will hear the jammed and watermarked message or just the watermarked message, depending on her location.
More details will be given later when we discuss the modelling of NSWJ using \Cref{fig:wbplsec_nswj3_csp}.

\textbf{Attack models and inference rules.}
In WBPLSec, Eve is usually passive.
She captures transmitted signals and tries to learn more knowledge from them.
In this paper, we also consider an active Eve who can send fake messages to legitimate agents.
The information she can derive and the messages she can convey are controlled by inference rules based on her current knowledge.
Such rules are defined in \Cref{table:inference} where $K$\changed[\Cn{1}{2}]{, a set of messages,} denotes Eve's current knowledge.
Rules are divided into two groups: \emph{breakdown} rules, \changed[\Cn{1}{2}]{where $\left(K \inferd m\right)$ denotes a message $m$ can be derived from $K$ using one of} member, unpairing, decryption, digital verify, watermarking, and jamming rules; and \emph{build-up} rules, \changed[\Cn{1}{2}]{where $\left(K\inferu m\right)$ denotes a message $m$ can be derived from $K$ using one of} member, pairing, encryption, digital sign, and watermarking rules.
We note that the Dolev-Yao cryptographic inference rules are also supported and included in the table to make our inference more general.

\begin{table*}[bt]
    \caption{Intruder message inference rules: breakdown and build-up rules.}
    \label{table:inference}\centering
\bgroup
\begin{tabular}{@{}l cl cp{3.3cm} cl cl @{}}
        \toprule
        Name & \phantom{a} & Id & \phantom{a} & Premises & \phantom{a}  & Break down & \phantom{a} & Build up \\
        \midrule
Member & & Mb & & $m \in K$ && $K \breakdown m$ && $K \buildup m$  \\
\midrule
Pairing && Pa && $m \in K \land m' \in K$ && && $ K \buildup (m,m')$  \\
Unpairing && Up && $(m,m') \in K$ && $K \breakdown m \land K \breakdown m'$ &&  \\
\midrule
Enc/Sign && Enc && $m \in K \land k \in K $ && && $K \vdash_{\Uparrow} \{m\}_k$ \\
Dec/Verify && Dec && $\{m\}_k \in K \land k^{-1} \in K$ && $K \vdash_{\Downarrow} m$ && \\
\midrule
\multirow{2}{3.5em}{Water-marking} && Wat1 && $\watm{m}{b} \in K$ && $K \breakdown m$ &&  \\
 && Wat2 &&$m \in K \land b \in K$ && && $ K \buildup \watm{m}{b}$  \\
\midrule
 Jamming && Jam && $\jam{\watm{m}{b}}{b'} \in K\land$ \newline $ b' \in K \land \exists x \bullet b' \cat x = b$ && $K \breakdown \watm{m}{b}$ && \\
        \bottomrule
\end{tabular}
\egroup
\vspace*{-2em}
\end{table*}

As with~\cite{Costa2023}, $\watm{m}{n}=\watm{m'}{n'}$ only if $m=m'$ and $n=n'$.
That means the same watermarked messages can be generated only if watermarking applies to the same message using the same bitmask. Also, $\jam{m}{0}=m$ denotes that jamming using an empty bitmask is like no jamming. Particularly, the jamming rule defines that if jamming a watermarked message using the prefixing bitmask $b'$ of the watermarking bitmask $b$, the watermarked message can be recovered when $b'$ is known. Here, we treat a bitmask as a sequence of bits and use the sequence concatenation $\cat$ to establish that $b'$ is the prefix of $b$. This corresponds to $M$ and $N_w$ in \ref{Step:3}.

Informally, if the intruder sits within the jamming range of a receiver $A$, she can only hear jammed watermarked messages. According to \ref{AS:3} and the inference rule Jam in \Cref{table:inference}, she cannot derive the watermarked messages from the learned messages. So, secrecy is preserved. However, if she sits outside the jamming range but still within the transmission range of the sender, she may still be able to hear watermarked messages. According to the Wat1 inference rule, she can derive clear messages from the watermarked messages. So, secrecy is violated. According to \ref{AS:3}, in both cases, message integrity is maintained because the intruder cannot forge a watermarked message without knowledge of its secret code. An active intruder can build up all messages from her knowledge using the build-up rules. And then she can fork or send these (possibly watermarked using her secret code) messages to the network. According to \ref{AS:4}, the messages are neither expected nor accepted by legitimate agents because the secret code is unknown to other agents. Indeed, there is little difference between passive and active attack models.

\textbf{Message data types.}
We define a generic message type which is parametric in the sizes of various entities in messages, such as agents, nonces, keys, bitmasks, and modulo exponent bases. We require these entities to be finite for the sake of the executable nature of animation. This generic type is achieved through a data type \isacode{fsnat}, defined below, for a finite set of natural numbers. Consequently, each entity (from a finite set of $n$ entities) can be numbered from 0 to $n-1$. 
\newcommand{\fsnat}{\isalink{\coderepofsnat\#L11}}
\begin{alltt} 
\isakwmaj{typedef} (\isakwmin{overloaded}) ('n::len) fsnat = "{0..<LENGTH('n)}::nat set" \(\fsnat\)
\end{alltt} 
The \isacode{fsnat} is a polymorphic type with one type variable \isacode{\textquotesingle{n}} of class \isacode{len}. The \isacode{len} class is equipped with a \isacode{LENGTH} function to get its size. So the new type \isacode{\textquotesingle{n} fsnat} is isomorphic to a finite set of natural numbers from 0 to the size of \isacode{\textquotesingle{n}} minus 1. This type comes with a pair of functions \isacode{nat\_of\_fsnat} and \isacode{fsnat\_of\_nat} to convert from \isacode{fsnat} to a natural number or from a natural number to a \isacode{fsnat}. To facilitate the construction of such a \isacode{\textquotesingle{n} fsnat} value, we define \isacode{nmk} (\isaref{\coderepofsnat\#L15})
 function to construct a \isacode{\textquotesingle{n} fsnat} value from its input natural number \isacode{x}. 
%
Then a variety of data types can be defined using \isacode{fsnat}. 
\newcommand{\dagent}{\isalink{\coderepomsg\#L14}}
\newcommand{\dnonce}{\isalink{\coderepomsg\#L57}}
\newcommand{\dexpg}{\isalink{\coderepomsg\#L59}}
\newcommand{\dkey}{\isalink{\coderepomsg\#L62}}
\newcommand{\dbitmask}{\isalink{\coderepomsg\#L118}}
\begin{alltt}
\isakwmaj{datatype} ('n::len) dagent = Agent (ag:"'n fsnat") |Intruder |Server\(\dagent\)
\isakwmaj{datatype} ('k::len,'s::len) dkey = Kp "fsnat['k]" | Ks "fsnat['s]" \(\dkey\)
\isakwmaj{datatype} ('bm, 'bl) dbitmask = Null | Bm  "'bm fsnat" "'bl fsnat" \(\dbitmask\)
\end{alltt}
Participants or agents are modelled using \isacode{\textquotesingle{n} dagent} where \isacode{\textquotesingle{n}} is the number of legitimate \isacode{Agent}s except an \isacode{Intruder} and a \isacode{Server}. In the definition, \isacode{ag} is a function to extract the agent number from an \isacode{Agent}. If such a function is applied to \isacode{Intruder} or \isacode{Server}, an exception will be raised in the generated code. Nonces (of type \isacode{dnonce}) and modulo exponent bases (\isacode{dexprg}) are just \isacode{fsnat} and their definitions are omitted here. Keys (of \isacode{dkey}) are either public by the constructor \isacode{Kp} or private by the constructor \isacode{Ks}. %

Bitmasks (of type \isacode{dbitmask}) contain an additional \isacode{Null} for an empty bitmask. The type has two type variables \isacode{\textquotesingle{bm}} and \isacode{\textquotesingle{bl}} to represent the number of bitstrings and the maximum length of a bitstring used for watermarking or jamming. 
Additionally, this data type is instantiated to partial \isacode{order}, so any two bitmasks can be compared using $\leq$. And two bitmasks \isacode{bm1} $\leq$ \isacode{bm2} if either 
\begin{enumerate*}[label=(\arabic*)]
    \item \isacode{bm1} is empty, or 
    \item both \isacode{bm1} and \isacode{bm2} are not empty (such as \isacode{Bm b1 l1} and \isacode{Bm b2 l2}), their bitstrings are equal (\isacode{b1=b2}), and their lengths are less than or equal (\isacode{l1}$\leq$\isacode{l2}). 
\end{enumerate*}
This is to implement the abstract prefix relation in \Cref{table:inference} to derive the watermarked message from its jammed counterpart.

Based on these types, a message type is defined below where only a few constructors are shown and the complete definition is given in \Cref{appendix:message:type}.
\newcommand{\dmsg}{\isalink{\coderepomsg\#L202}}
\begin{alltt}
\isakwmaj{datatype} ('a, 'n, 'k, 's, 'g, 'bm, 'bl::len) dmsg = \ldots \(\dmsg\)
  | MBitm   (mbm:"('bm, 'bl) dbitmask")                \(\hfill\bitmask{\_ }{\_}\) 
  | MWat    (mwm:"T dmsg") (mwb:"T dmsg")\(\hfill\mwat{\_}{\_}\) 
  | MJam    (mjm:"T dmsg") (mjb:"T dmsg")\(\hfill\mjam{\_}{\_}\) 
\end{alltt}
%
All messages for communication on channels are of the polymorphic type \isacode{dmsg} with 6 type variables to represent the numbers of agents (\isacode{\textquotesingle{a}}), nonces (\isacode{\textquotesingle{n}}), public keys (\isacode{\textquotesingle{k}}), private keys (\isacode{\textquotesingle{s}}), modulo exponent bases (\isacode{\textquotesingle{g}}), and bitmasks (\isacode{\textquotesingle{b}}). \isacode{T} in the definition above is a shorthand for \isacode{(\textquotesingle{a}, \textquotesingle{n}, \textquotesingle{k}, \textquotesingle{s}, \textquotesingle{g}, \textquotesingle{bm}, \textquotesingle{bl}::len)}. Messages can be agent's identities of type \isacode{dagent} (with the constructor \isacode{MAg}), nonces (with \isacode{MNon}), public and private keys (with \isacode{MK}), pairing of two messages (with \isacode{MPair}), 
modulo exponent bases (with \isacode{MExpg}), modulo exponentiation (with \isacode{MModExp}), bitmasks (with \isacode{MBitm}), watermarking (with \isacode{MWat}), and jamming (with \isacode{MJam}). %
To simplify and shortening formulas, we introduce the notations $\pairmsg{m_1}{m_2}$, 
$\modbase{n}$, $\modexp{m}{e}$, $\bitmask{n}{l}$, $\mwat{m}{b}$, and $\mjam{m}{b}$ to denote them respectively.

\textbf{Message inferences.} In Isabelle, we define a function \isacode{breakm} (\isaref{\coderepomsg\#L492}) for the breakdown rules in \Cref{table:inference} to derive a list of messages from a given list for the sake of implementation in Haskell. The \isacode{breakm} is defined on another function \isacode{break\_lst} (\isaref{\coderepomsg\#L366}). We omit the details of these functions here for simplicity.

For the build-up rules, \cite{Ye2024a} uses \isacode{buildm} to construct all possible messages by these rules. The number of these messages is usually huge. It is time-consuming and space-consuming, and has a significant impact on performance. To address this issue, we propose a solution that considers only the possible acceptable messages from all legitimate agents and disregards other unacceptable messages. This is a reasonable assumption because the transmission of these unacceptable messages will not change the behaviour of all legitimate agents. 

For the approach, we first define a function \isacode{buildable m ms} (\isaref{\coderepomsg\#L703}) to return whether a message \isacode{m} can be constructed from a set \isacode{ms} of messages using the build-up rules.
Then \isacode{filter\_buildable xs ms} (\isaref{\coderepomsg\#L726}) returns a list of messages, of which each message is from \isacode{xs} and buildable from \isacode{ms}. 
This improves the inference performance dramatically in animation.

Generally, when the intruder hears a new message, she attempts to break it down using her existing knowledge and updates it accordingly. Then, if modelling an active attack, she will construct all buildable messages from her up-to-date knowledge using the \isacode{filter\_buildable} function and send them to the network.

\textbf{Signals.} We declare signals to specify properties. 
The definition \isacode{dsig} (\isaref{\coderepomsg\#L840}) is the same as~\cite{Ye2024a} and omitted here.

\textbf{Channels.} 
Channels for communication are grouped in a definition \isacode{chan} below, where seven channels (with their corresponding message types) are declared.
\newcommand{\chan}{\isalink{\coderepomsg\#L844}}
\begin{alltt}
\isakwmaj{datatype} T chan =\(\chan\)
  env   :: "'a dagent\(\cross\)'a dagent"
  send  :: "'a dagent\(\cross\)'a dagent\(\cross\)'a dagent\(\cross\)T dmsg"
  recv  :: "'a dagent\(\cross\)'a dagent\(\cross\)'a dagent\(\cross\)T dmsg"
  cjam  :: "T dmsg"    sig   :: "('a, 'n) dsig"   
  leak  :: "T dmsg"    terminate :: "unit"
\end{alltt}
We extend the traditional channels \isacode{send} and \isacode{recv} in~\cite{Ye2024a} with an additional component such as \isacode{M} in \isacode{send!(A, M, B, m)} to denote the medium through which the message is sent or received.
If \isacode{M} is the \isacode{Intruder}, then it denotes a message through a public and insecure channel.
Otherwise, it implies the use of a private channel (so the intruder cannot hear such a message).
With this additional component, we can model both public and private channels.
And \isacode{hear} and \isacode{fake} (or \isacode{relay}) are just synonyms to \isacode{send} and \isacode{recv}.
The channel \isacode{cjam} relates to jamming communications. The channels \isacode{sig}, \isacode{leak}, and \isacode{terminate} are used to signal the stages which the protocol is on. Eventually, they are used to specifying security protocols. For example, secrecy is specified using \isacode{leak} events and authenticity is specified through signals \isacode{StartProt} and \isacode{EndProt} on channel \isacode{sig}. The \isacode{terminate} event indicates a successful completion of a protocol run.


In brief, our formalization establishes structured inference rules to determine what an intruder can know or do, while models, watermarking, and jamming restrict their capabilities through physical-layer security techniques.

\section{Needham-Schroeder and Diffie-Hellman Protocols}
\label{sec:nsdh}


The Needham-Schroeder protocol~\cite{Needham1978} is used to establish mutual authentication between two parties (such as Alice and Bob) over an insecure or public network, utilizing either symmetric or asymmetric encryption.
In this paper, we consider a variant of NSPK, referred to as NSWJ here, which utilizes watermarking and jamming (instead of asymmetric encryption) to achieve secrecy and authenticity, similar to~\cite {Costa2023}. 
We illustrate the message exchange between Alice and Bob for both NSPK and NSWJ below, where the asymmetric encryption in NSPK is replaced by watermarking in NSWJ. 
\begin{align*}
    \begin{array}{lcl}
    \qquad\qquad NSPK & \qquad\qquad & \qquad\qquad NSWJ \\ 
    A \to B: \aenc{(na, Alice)}{\pk{k}{B}} & & A \to B: \mwat{(na, Alice)}{bA_w} \\ 
    B \to A: \aenc{(na, nb)}{\pk{k}{A}} &  &B \to A: \mwat{(na, nb)}{bB_w} \\ 
    A \to B: \aenc{nb}{\pk{k}{B}} &  &A \to B: \mwat{nb}{bA_w}
    \end{array}
\end{align*}
\begin{figure}[bt]
    \centering
    \includegraphics[width=0.65\columnwidth]{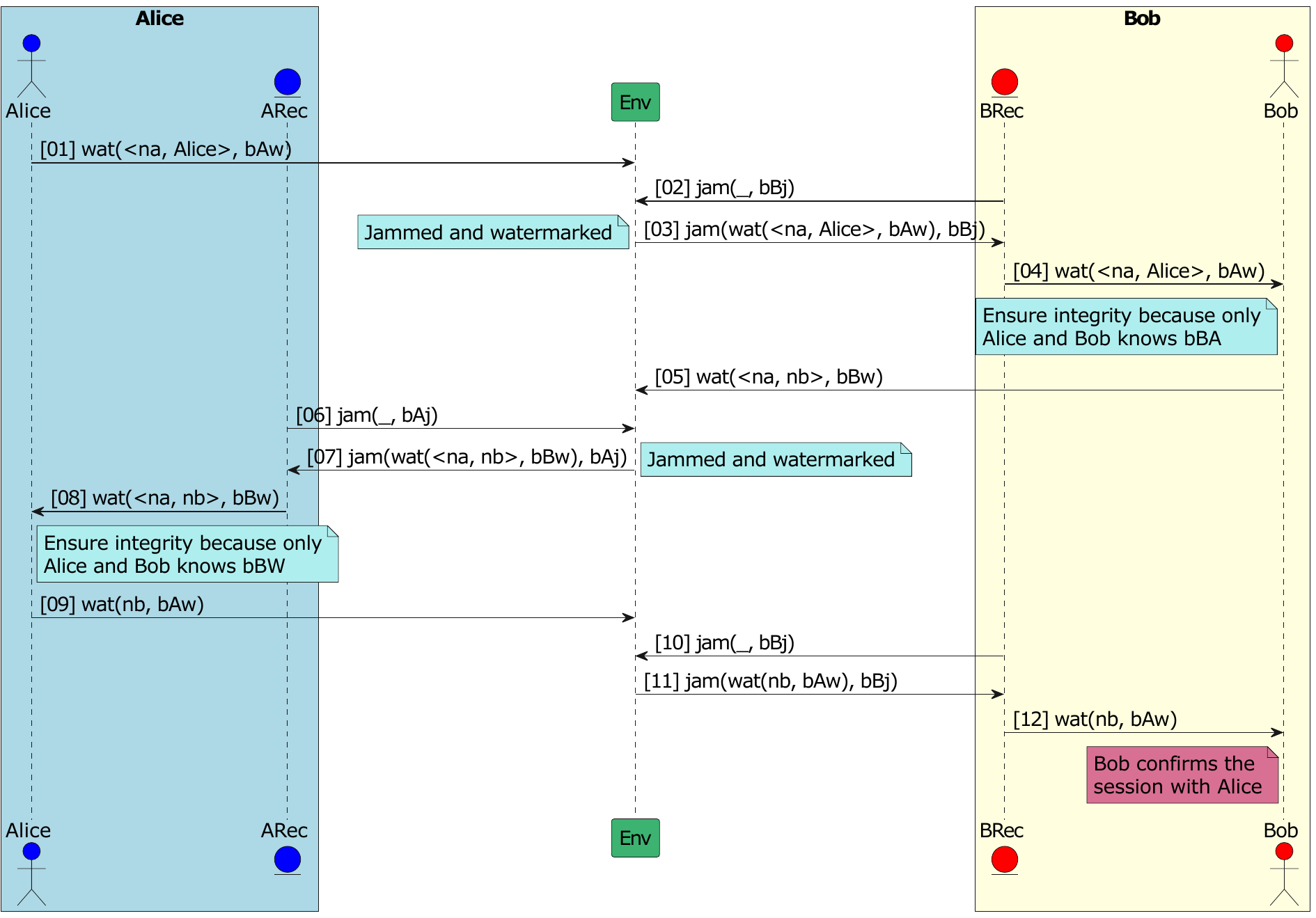}
    \caption{The messages between Alice and Bob for NSWJ.}
    \label{fig:wbplsec_nswj3}
    \vspace*{-1em}
\end{figure}
We further illustrate the detailed message exchange in NSWJ in \Cref{fig:wbplsec_nswj3}, where ARec and BRec represent the receivers of Alice and Bob, respectively.
Alice starts the session by sending a watermarked message (01) to the environment Env (i.e., the secure region) containing a nonce {\tt na} and her identity, using her watermarking bitmask.
BRec starts to jam (02) Alice's watermarked message using Bob's jamming bitmask and receives the (03) jammed watermarked message from the environment.
BRec can recover the watermarked message because BRec knows the jamming code and sends it (04) to Bob.
Bob can confirm this message is from Alice because it is watermarked using Alice's bitmask.
Afterwards, Bob sends (05) back Alice's nonce {\tt na} and his nonce {\tt nb} to Env.
Similarly, ARec will jam (06) the message, receive (07) the jammed and watermarked message, recover the watermarked message from Bob, and send (08) it to Alice.
Alice confirms the integrity of the message and sends (09) {\tt nb} back to Bob.
BRec will jam (10), receive (11), recover, and send (12) it to Bob.
Finally, Bob can confirm he has established a secure session with Alice if Eve is only within the jamming range of Alice and Bob.


%


%

\begin{figure*}[tb]
    \centering
    \includegraphics[width=0.85\textwidth]{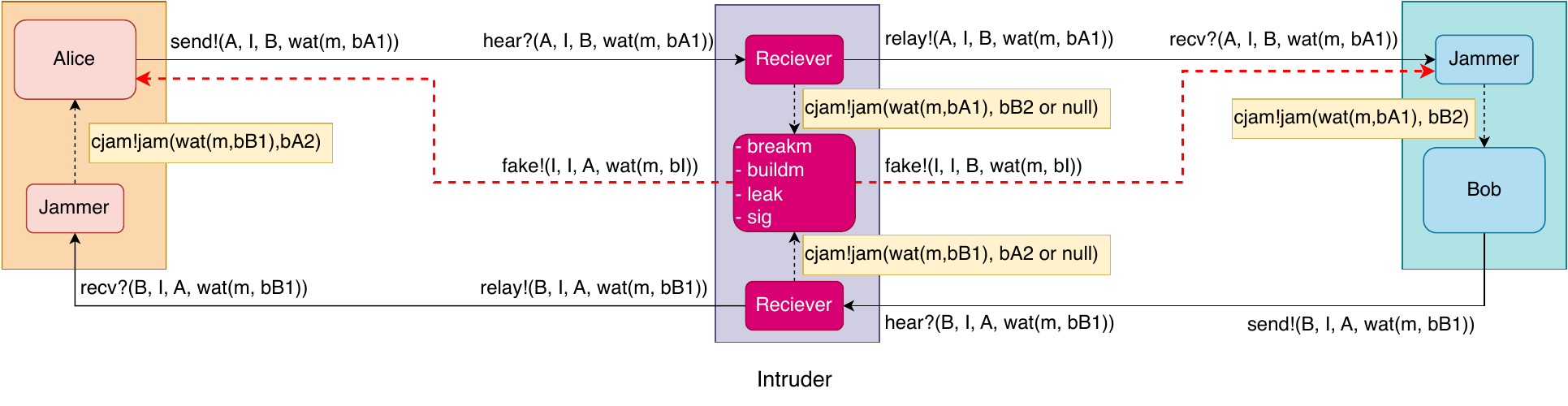}
    \caption{The modelling of NSWJ in ITree-based CSP.}
    \label{fig:wbplsec_nswj3_csp}
    \vspace*{-2em}
\end{figure*}

\Cref{fig:wbplsec_nswj3_csp} schematically depicts the modelling of NSWJ in ITrCSP.
We model Alice, Bob, and the intruder as three processes.
Both Alice and Bob contain a jamming receiver (or jammer), and the intruder includes a receiver (\Cref{fig:wbplsec_nswj3_csp} splits the receiver into two elements for better visualisation of connections).
Alice and Bob \cspcode{send} messages to the intruder and receive messages on the \cspcode{recv} channel from the intruder.
The intruder \cspcode{hear}s messages that are sent by Alice or Bob and can \cspcode{fake} messages (if he is an active attacker) that Alice and Bob will receive.
The receiver of the intruder will relay the received watermarked messages to intended agents (such as Bob), and jam the messages using the bitmasks (such as \cspcode{bB2}) of their intended agents (or a \cspcode{Null} bitmask) if he is in the jamming range (or out of the range) and send the jammed messages to the intruder through the \cspcode{cjam} channel.
While the intended agents receive the relayed watermarked messages, the jammer jams the messages using his jamming bitmask (such as \cspcode{bB2}) and sends the messages to the agents (such as Bob).
This way, we can flexibly model the location of the intruder, and his location will not affect the message Bob receives, because the intruder's receiver will jam the watermarked messages accordingly based on his location.
We note that the \cspcode{cjam} channel is represented by a dashed line, indicating that jamming is an internal behaviour and will be hidden from the modelled protocol in CSP.

The main body of the intruder, represented as a box in the middle, describes its main behaviour: breaking down heard messages, building up new fake messages, and sending them to the network (if it is active), leaking secrets if the secrets are within its knowledge, or dealing with signals.

To model the protocol in Isabelle using ITrCSP, we first define a \emph{configuration} of this protocol, which includes the instantiation of message types as follows.
\newcommand{\nswjconfig}{\isalink{\codereponswjconf\#L5}}
\newcommand{\nswjconfigagent}{\isalink{\codereponswjconf\#L15}}
\begin{alltt} 
\isakwmaj{type_synonym} max_agents = 2; max_nonces = 4; max_bm = 3; \(\nswjconfig\)
    max_bm_len = 2; max_pks = 1; max_sks = 1; max_expg = 1
\isakwmaj{type_synonym} dagent = "max_agents dagent"\(\nswjconfigagent\)
\end{alltt}
First, we define type synonyms, such as \isacode{max\_agents}, as a numeral type 2 and use these synonyms to instantiate polymorphic data types, \isacode{dagent}, etc. We give the same names to type synonyms of these concrete types. When we refer to these types in the rest of the paper, we mean the specific types mentioned in this section. In this protocol, we define three bitmasks, each with a length of up to 2 (\isacode{max\_bm\_len}). The protocol is configured below.
\newcommand{\nswjconfigalice}{\isalink{\codereponswjconf\#L21}}
\begin{alltt}
\isakwmaj{abbreviation} "Alice \(\equiv\) Agent (nmk 0) :: dagent"\(\nswjconfigalice\)
\isakwmaj{abbreviation} "Bob \(\equiv\) Agent (nmk 1) :: dagent"
\isakwmaj{abbreviation} "nonmap \(\equiv\) \(\{\)Alice\({}\mapsto{}\) nmk 0, Bob\({}\mapsto{}\)nmk 1, Intruder\({}\mapsto{}\)nmk 2\(\}\)"
\isakwmaj{abbreviation} "bmmap \(\equiv\) \(\{\)Alice\({}\mapsto{}\)Bm (nmk 0) (nmk 1), 
  Bob\({}\mapsto{}\)Bm (nmk 1) (nmk 1), Intruder\({}\mapsto{}\)Bm (nmk 2) (nmk 1)\(\}\)"
\end{alltt}
Such a configuration maps the legitimate agents Alice and Bob to the agent, nonce, and bitmask indices, where the constructors like \cspcode{nmk} are used to make an instance of \cspcode{fsnat}. The location of an intruder is defined in terms of jamming ranges and configured using the map below. 
\newcommand{\nswjconfigdeve}{\isalink{\codereponswjconf\#L67}}
\newcommand{\nswjconfigrange}{\isalink{\codereponswjconf\#L94}}
\begin{alltt}
\isakwmaj{datatype} deve = Eve1 | Eve2 | Eve3 | Eve4\(\nswjconfigdeve\)
\isakwmaj{abbreviation} "evemap eve \(\equiv\) case eve of \(\nswjconfigrange\)
    Eve1 \(\Rightarrow\) \(\{\)Alice\({}\mapsto{}\)T, Bob\({}\mapsto{}\)F\(\}\) | Eve2 \(\Rightarrow\) \(\{\)Alice\({}\mapsto{}\)F, Bob\({}\mapsto{}\)T\(\}\) | 
    Eve3 \(\Rightarrow\) \(\{\)Alice\({}\mapsto{}\)T, Bob\({}\mapsto{}\)T\(\}\) | Eve4 \(\Rightarrow\) \(\{\)Alice\({}\mapsto{}\)F, Bob\({}\mapsto{}\)F\(\}\) 
\end{alltt}
The \cspcode{evemap} configures the location (among four, see \Cref{fig:wbplsec}) of the intruder (\isacode{eve}) based on where the intruder is within the jamming range of Alice and Bob (where \isacode{T} for True and \isacode{F} for False). 
Then we use these configurations to define the jamming process (\isacode{jamming} \isaref{\codereponswj\#L44}) for Alice and Bob, the jamming process (\isacode{jammingI} \isaref{\codereponswj\#L55}) for the intruder, the process (\isacode{PAlice} \isaref{\codereponswj\#L133}) for Alice, the process (\isacode{PBob} \isaref{\codereponswj\#L207}) for Bob, and the process (\isacode{PIntruder} \isaref{\codereponswj\#L296}) for Intruder where her initial knowledge is taken into account. We discuss the definition of \isacode{PAlice} more below and omit the details of other processes here for simplicity.
\begin{align*}
    PAlice \defs & \left(\left(
    \begin{array}{l}
        Initiator(\_) \parallel_{jevents} jamming(\_)
    \end{array}\right) \hide jevents\right) 
    \exceptp{}{\{Terminate\}}{skip} 
\end{align*}
\isacode{PAlice} is a parallel composition of an \isacode{Initiator} (see \cite{Ye2024a} for details) for the behaviour of Alice, and \isacode{jamming} over the jamming events (\isacode{jevents}) with these events hidden. This process can be terminated using the CSP exception operator over the \isacode{Terminate} channel.
Finally, the protocol NSWJ is composed of the processes for Alice, Bob, and the intruder in response to appropriate events.
\newcommand{\pnspk}{\isalink{\codereponspk\#L263}}
\begin{align*}
    \isacode{NSWJ} \defs \left({PAlice} \parallel_{TermEvent} {PBob}\right)\parallel_{ABIEvents} {PIntruder} \isalink{\codereponswj\#L345}
\end{align*}
This is the process used to generate an animator for the protocol.

The Diffie-Hellman (DH)~\cite{Diffie1976} protocol aims to establish a shared secret between two agents using agreed-upon information over an insecure network. It is commonly used to derive new session keys.
DH is based on modular exponentiation $g^a \mod p$ where $g$ is the base, $a$ is the power or exponent, and $p$ is the modulus. 
In \cite{Ye2024a}, we model the DH with three messages:  
\begin{enumerate*}[label={(\arabic*)}]
        \item {$ A \to B: g^{na} $} and $B$ computes $k_B = g^{na^{nb}}$,
        \item $ B \to A: g^{nb}$ and $A$ computes $k_A = g^{nb^{na}}$,
        \item $ A \to B: \enc{t}{k_A}$ where $t$ is a secret and the superscript $s$ denotes a symmetric encryption, and $B$ decrypts it using $k_B$ to get $t$.
\end{enumerate*}
Using sound animation, it is easy to demonstrate a man-in-the-middle attack where secrecy is not preserved.

Next, we consider a variant of DH, called DHWJ, which uses watermarking and jamming. It is composed of four messages: 
\begin{enumerate*}[label={(\arabic*)}]
    \item {$ A \to B: \mwat{g^{na}}{bA_w} $} and $B$ computes $k_B = g^{na^{nb}}$,
    \item $ B \to A: \mwat{g^{nb}}{bB_w}$ and $A$ computes $k_A = g^{nb^{na}}$,
    \item $ A \to B: \mwat{\enc{t}{k_A}}{bA_w}$ and $B$ decrypts it using $k_B$ to get $t$,
    \item $ B \to A: \mwat{\enc{t}{k_B}}{bB_w}$ and $A$ decrypts it to confirm $B$ knows $t$. 
\end{enumerate*}
The message exchange and protocol configuration (\isaref{\coderepodhwj}) are similar to those of NSWJ and are omitted here for simplicity.

\section{Verification and Evaluation}
\label{sec:verification}

In Isabelle, we can automatically generate the Haskell code for the NSWJ and DHWJ protocols.
To facilitate the automatic or manual exploration of the design to check secrecy and authenticity, we developed a new web interface\footnote{The source code is available at \url{https://github.com/RandallYe/Animation_of_Security_Protocols/tree/master/animation-web-ui}} in addition to the terminal interface (as presented in~\cite{Ye2024a}) to directly visualise sequence diagrams of animation and counterexamples (attacks), as illustrated in \Cref{sec:web_interface}. %
\changed[\Cn{1}{3}]{The verification algorithm underneath interfaces is the depth-first exploration of the event space up to specified steps for a protocol design. The terminal interface provides users with manual exploration, automatically exhaustive or random reachability check in terms of trace properties, or feasibility check of a specified sequence of events or trace.
Random exploration and feasibility check are not yet implemented in the web interface. It, however,} supports sessions and enables designers to (simultaneously) manually explore or exhaustively check reachability and view counterexamples visually in sequence diagrams, as shown in \Cref{fig:wbplsec_nswj3}. 
Graphical visualisation of protocols as sequence diagrams has been demonstrated in~\cite{Metere2022}; however, their approach is design-oriented rather than analysis-oriented, and their diagrams are manually created, whereas our diagrams are dynamically updated.
We could not extend it, as manually created diagrams lack the precise execution semantics needed for automated visualisation and attack simulation.

While the web interface provides no more verification features than the terminal interface functionally, the diagrammatic view of counterexamples helps designers easily follow their interactions with animators and identify problems. Additionally, verification is more easily configured through the web interface, allowing for the selection of intruder location and properties to check. For both protocols, the web interface will enable users to select which location of Eve to animate and verify, manually animate the protocol, automatically verify secrecy and correspondence properties such as authentication, or combine manual exploration with an automatic check for user-guided verification. 
More details about the web interface can be found in~\Cref{sec:web_interface}.

\begin{table*}[bt]
    \caption{Comparison of verification results: NSPK vs. NSWJ, and DHWJ vs. DH in the presence of an active intruder, except the specific NSWJ$^\dagger$ for a passive intruder.}
    \label{table:result:NSWJ}\centering
\bgroup
\begin{tabular}{@{}l c c c lccccl c lccccl c c c lcccc@{}}
        \toprule
        \multirow{2}{*}{Properties} & \tablecolsep &  \multirow{2}{*}{NSPK} & &  \multicolumn{6}{c}{NSWJ} & \tablecolsep &  \multicolumn{6}{c}{NSWJ Passive$^\dagger$} & &  \multirow{2}{*}{DH} & \tablecolsep &  \multicolumn{5}{c}{DHWJ}\\
        \cline {6-9}
        \cline {13-16}
        \cline {22-25}
         & & & && E1 &E2 &  E3 &  E4 &&  & &E1 &  E2 &  E3 &  E4 &&& && &E1 &  E2 &  E3 &  E4 \\ 
        \midrule
        Secrecy                 && $\circ$ && &$\circ$ &$\circ$ & $\bullet$ & $\circ$ && &&$\circ$ & $\circ$ & $\bullet$ & $\circ$ &&& $\circ$ &&& $\circ$ & $\circ$ & $\bullet$ & $\circ$ \\ 
        Authenticity for Alice  && $\bullet$ && &$\bullet$ &$\bullet$ & $\bullet$ & $\bullet$ && &&$\bullet$ & $\bullet$ & $\bullet$ & $\bullet$ &&& $\circ$ &&& $\bullet$ & $\bullet$ & $\bullet$ & $\bullet$ \\ 
        Authenticity for Bob     && $\circ$ && &$\bullet$ &$\bullet$ & $\bullet$ & $\bullet$ && &&$\bullet$ & $\bullet$ & $\bullet$ & $\bullet$ &&& $\circ$ &&& $\bullet$ & $\bullet$ & $\bullet$ & $\bullet$ \\ 
        \bottomrule
\end{tabular}
\egroup
\vspace*{-1em}
\end{table*}

Using the interfaces, we analysed NSWJ and DHWJ, along with their cryptographic counterparts, NSPK3 and DH, in terms of secrecy and authenticity. \Cref{table:result:NSWJ} shows verification results in the presence of an active intruder and NSWJ in the presence of a passive intruder. The results of NSPK and DH are already discussed in~\cite{Ye2024a}. It is not surprising that secrecy is only satisfied in the scenario of Eve3 (E3 in the table) because Eve can only hear jammed messages from both Alice and Bob (so she cannot derive clear messages). Authenticity for both Alice and Bob is preserved when using WBPLSec due to the watermarking feature. 
We note that DH is not an authentication protocol and usually other mechanisms such as digital signature, hash, trusted certificate authority (CA), and message authentication code (MAC) are required to achieve both secrecy and authenticity such as in the Unified model and MQV~\cite{BlakeWilson1999}, SIGMA~\cite{Krawczyk2003} and HMQV~\cite{Krawczyk2005}. With PLS, we demonstrate that both properties can be achieved without such a mechanism, regardless of Eve's location.

While the ProVerif analysis in~\cite{Costa2023} uses an active intruder, we consider both active and passive intruders. 
Our animation shows that whether the intruder is active or passive makes no difference in terms of secrecy and authenticity.
If a property is held in the passive model, it is also held in the active model and vice versa.
This is related to \ref{AS:4} where neither Alice nor Bob knows the Intruder's watermarking bitmask, and therefore neither Alice nor Bob will accept the watermarked (inferred) message from the Intruder. 
This should not be surprising because watermarking can ensure message integrity.

Although the result of NSWJ in~\cite{Costa2023} is correct and consistent with ours, we argue that their speculation ``WBPLSec, in which the receiver, besides the sender, also plays an active role in the communication, so keeping authenticity between the enrolled parties'' is not accurate.
It is recognized that the real problem with NSPK is its second message $\pairmsg{na}{nb}$, sent by Bob to Alice, which contains no information about Bob's identity (authenticity is not guaranteed).
This permits a man-in-the-middle to forward such a message to Alice, making her believe it originated from the intruder (while Bob generated it).
To support that, the fix~\cite{Lowe1996} appends Bob's identity to the second message.
We conclude that NSWJ ensures authenticity not due to jamming, but instead to watermarking, thanks to its implicit integrity guarantee.
This evidence suggests that our formalism can enhance the understanding of protocols and their security mechanisms for designers.

\changed[\Cn{2}{5}, \Cn{2}{8}]{
    To evaluate performance, we ran all the verification on the same MacBook Pro laptop. For both the original NSPK and its corrected version~\cite{Lowe1996}, it took about 1 second (s) in Tamarin and about 200 milliseconds (ms) in ProVerif to verify all secrecy and authenticity properties. It took about 4 seconds in our tool for each property. For the NSWJ version in~\cite{Costa2023}, ProVerif uses 186 ms to prove four secrecy properties when considering Eve3. For the same protocol, our tool uses less than 1 second to verify each secrecy or authenticity property when considering the same eavesdropper location Eve3. However, it will take about 7 seconds to verify each secrecy property when considering other eavesdropper locations and about 9 (or 33) seconds to verify an authenticity property when considering Eve1 or Eve2 (or Eve4). So, it takes longer for our tool to verify properties, which is due to the executable nature of our approach to achieve accessibility for designers. The verification of the Diffie-Hellman protocols, including DH and DHWJ is similar.
}
In a nutshell, we successfully address all research questions. 
\changed[\Cn{1}{4}]{
    We extended our framework to support both the Dolev-Yao and PLS attack models, and verified two PLS-based security protocols (\ref{RQ1:pls}). 
    We generalised a message theory to be generic. So security protocols using different primitives and attack models can be modelled and verified in the same framework (\ref{RQ2:general}).
    We proposed DHWJ. With a thorough analysis of NSWJ and DHWJ using our approach, we found that authenticity is due to the watermarking (as we expect) and is preserved in all scenarios, even when secrecy is compromised. That is one benefit that PLS can bring (\ref{RQ3:cases}).
    We developed a web interface and used it to verify both the cryptographic-based and PLS-based WJ and DH protocols. Indeed, this can be carried out by a wide range of stakeholders (\ref{RQ4:accessibility}).
}

\section{Conclusion and Future Work}
\label{sec:concl}
This paper presents a formal verification framework for Physical Layer Security protocols based on sound animation with graphical interface to improve accessibility.
Our work builds upon the formalism by~\cite{Ye2024a}, based on ITrCSP and implemented in Isabelle/HOL. We further extend its generality by supporting both cryptographic and PLS protocols, by improving its intruder's message inference rules, and by introducing a new web interface.  
We apply our framework to model and verify the WBPLSec-based Needham-Schroeder protocol proposed in~\cite{Costa2023}. We further integrate watermarking and jamming techniques into the Diffie-Hellman protocol, and our analysis shows that the proposed protocol preserves both secrecy and authenticity when a new session key is derived.

\changed[\Cn{2}{9}]{Our work has limitations. We verify protocols through the exploration of event trees using animators. The event trees could be too big to be exhaustively explored for complex security protocols. For the same reason, our tool needs a longer time to verify these protocols. So, scalability is still a challenge. }
Our current approach reduces the intervention from formal experts, which is still needed to model security protocols in Isabelle initially. They need to write CSP processes for each legitimate agent and a process for the intruder, consider what messages these processes can send and receive, and then compose the processes together through corresponding events. 
With significant additional effort, improvements would require to
\begin{enumerate*}[label=(\arabic*)]
    \item design or utilise a domain-specific language (DSL) by designers to capture their protocols, properties, and attack models;
    \item automatically transform the DSL models to ITrCSP in Isabelle; and 
    \item automatically generate animators.
\end{enumerate*}

Following \cite{Costa2023}, we consider a simplified scenario where the jamming signal remains consistent within a given distance, and no signal is received beyond this distance. In reality, the jamming signal and transmission signal are degraded as the distance increases. We use this assumption to simplify scenarios in terms of the eavesdropper's location: either within the jamming range or outside the range. One of our future projects is to consider more complex scenarios.

Future work will also focus on verifying more cryptographic protocols, such as the Mesh Commissioning Protocol (MeshCoP) for the Thread network and variants of Diffie-Hellman protocols, such as the STS, ISO, and SIGMA protocols~\cite{Krawczyk2003}, and the Ephemeral Diffie-Hellman Over COSE\footnote{\url{https://datatracker.ietf.org/doc/rfc9528/}.}, and other PLS protocols. 

We also plan to deploy our verified case studies on a public website, allowing researchers, engineers, or students in security to explore and verify protocols interactively using our sound animation. 

\smallskip
\noindent \textbf{Acknowledgements.}  

The EPSRC and DSIT support this work through the Communications Hub for Empowering Distributed Cloud Computing Applications and Research (CHEDDAR) under grants EP/X040518/1, EP/Y037421/1 and EP/Y019229/1.
We thank Michele Sevegnani, Yue Gu, and Sana Hafeez from the University of Glasgow, Dalal Alrajeh, and Zhi Zhang from Imperial College London for their helpful comments on an early version of the manuscript.

\bibliographystyle{splncs04}
\bibliography{main,pls}

\appendix

\section{Soundness illustration}
\label{appendix:soundness}
\begin{figure}[t]
    \centering
    \includegraphics[width=\textwidth]{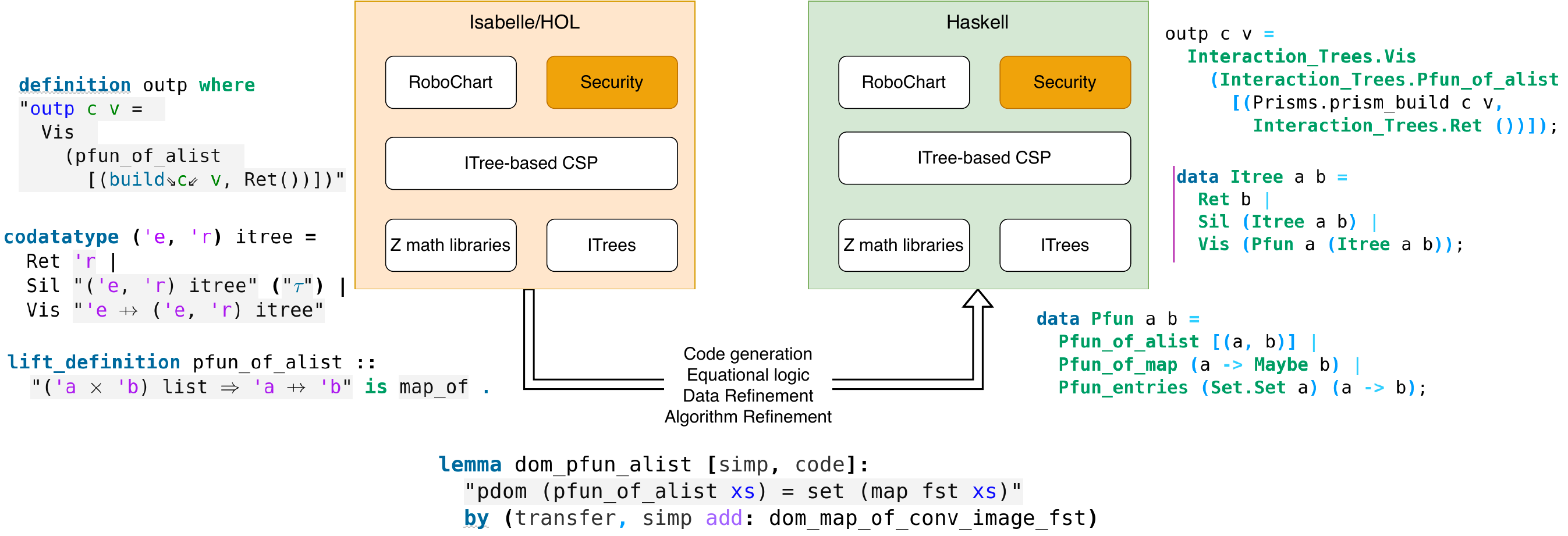}
    \caption{An example to illustrate how a CSP operator \cspcode{outp} (output a value \cspcode{v} on channel \cspcode{c}) is implemented on ITrees using associative lists (\cspcode{alist}) to represent partial functions (\cspcode{pfun}), and their counterparts in generated Haskell code where \cspcode{dom\_pfun\_alist} is a proved lemma in Isabelle/HOL and based on equational logic to replace its left-hand side by its right-hand side in code generation to preserve semantics. We say our approach from CSP specification to the resultant Haskell code forms an unbroken link.}
    \label{fig:sound_animation}
\end{figure}

\Cref{fig:sound_animation} shows how the definition of a CSP operator \cspcode{outp} in ITrees, on the left of the diagram, is linked to the final Haskell function \cspcode{outp} on the right of the diagram. This link is automated in Isabelle/HOL.

\section{Message type}
\label{appendix:message:type}

The message type in our framework is defined below.
\begin{alltt}
\isakwmaj{datatype} ('a, 'n, 'k, 's, 'g, 'bm, 'bl::len) dmsg = \(\dmsg\)
    MAg (ma:"'a dagent") 
  | MNon (mn:"'n dnonce") 
  | MK (mk:"('k, 's) dkey") 
  | MPair (mc1:"T dmsg") (mc2:"T dmsg")\(\hfill\pairmsg{\_}{\_}\) 
  | MExpg   (eg:"'g dexpg")                   \(\hfill\modbase{\_}\) 
  | MModExp (mmem:"T dmsg") (mmek:"T dmsg")\(\hfill\modexp{\_}{\_} or \mbox{\_\^{}\_}\) 
  | MBitm   (mbm:"('bm, 'bl) dbitmask")                \(\hfill\bitmask{\_ }{\_}\) 
  | MWat    (mwm:"T dmsg") (mwb:"T dmsg")\(\hfill\mwat{\_}{\_}\) 
  | MJam    (mjm:"T dmsg") (mjb:"T dmsg")\(\hfill\mjam{\_}{\_}\) 
  | MSEnc   (msem:"T dmsg") (msek:"T dmsg")  \(\hfill\ssec{\_}{\_}\) 
  | MAEnc   (mem:"T dmsg") (mek:"T dmsg")   \(\hfill\asec{\_}{\_}\) 
  | MSig    (msd:"T dmsg") (msk:"T dmsg")    \(\hfill\dsig{\_}{\_}\) 
\end{alltt}
 
\section{Web interface}
\label{sec:web_interface}
\begin{figure*}[tb]
    \centering
    \includegraphics[width=0.8\textwidth]{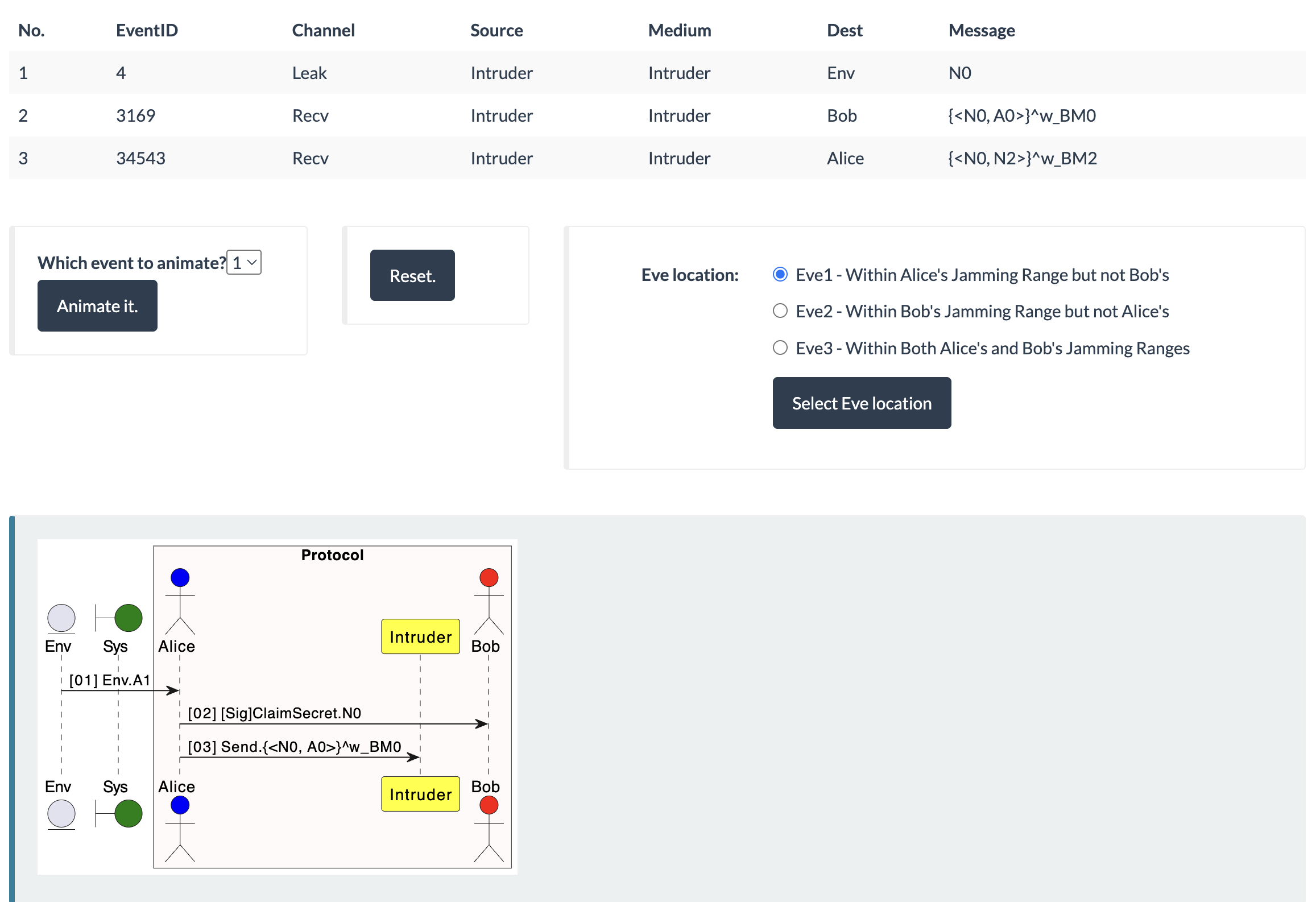}
    \caption{Manual animation: choose one numbered event to animate, and the interaction history is dynamically displayed in the sequence diagram.}
    \label{fig:nswj3_web_man}
\end{figure*}

\begin{figure}[ht!]
    \subfloat[][Configuration interface]{
  \begin{minipage}[b]{0.55\linewidth}
    \begin{center}
        \includegraphics[width=1.0\textwidth]{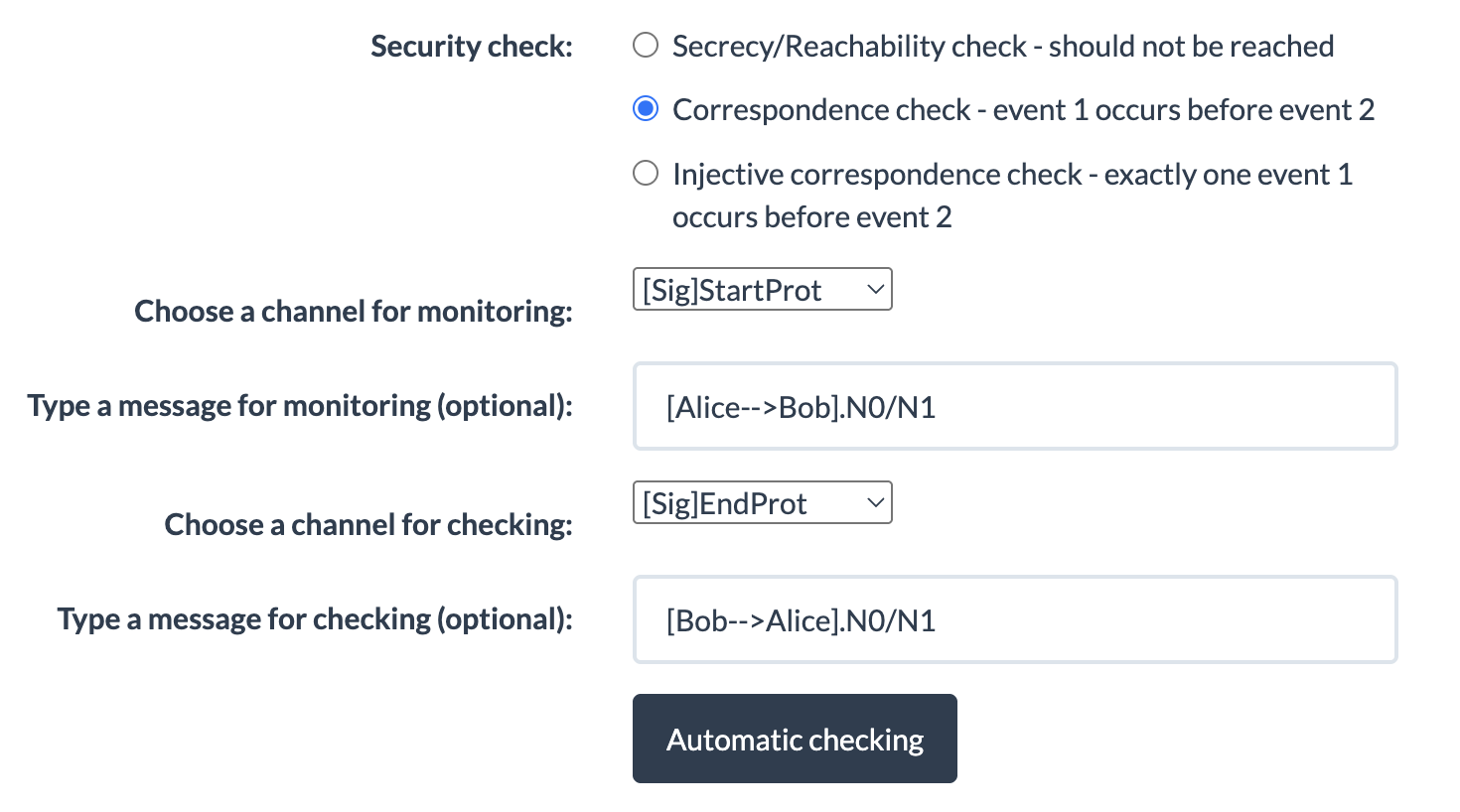}
    \end{center}
  \end{minipage}
    \label{fig:nswj3_web_auto}
}
\subfloat[][Secrecy counterexample]{
  \begin{minipage}[b]{0.45\linewidth}
    \begin{center}
        \includegraphics[width=0.7\textwidth]{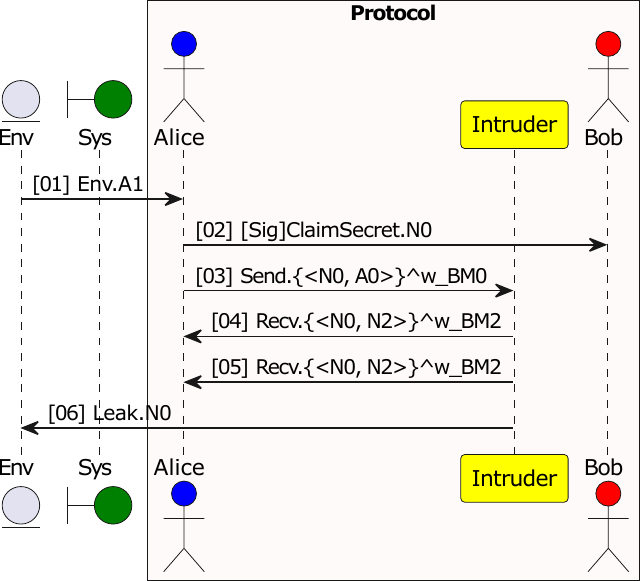}
    \end{center}
  \end{minipage}
  \label{fig:nswj3_secrecy_out_range_bob}
}
    \caption{Automatic reachability and correspondence checking for secrecy and authenticity.}
    \label{fig:nswj3_web_auto:check}
\end{figure}

We illustrate our web interface in \Cref{fig:nswj3_web_man} and \Cref{fig:nswj3_web_auto}.

\Cref{fig:nswj3_web_man} shows the manual exploration interface for NSWJ3, where the upper part of the table displays current enabled and available events for users to choose for animation. These events are numbered from 1. In the figure, there are three enabling events from 1 to 3. Each event has a unique identity and is shown in the second column. Additionally, the channel name, source, medium, target agents, and message body of an event are displayed. While Env is not a defined agent in our protocol, we use it to denote the environment of protocols or the users within them. For example, the second event is a message watermarked by Alice's bitmask (BM0) from a paired message composed of the nonce (N0) and the identity (A0) of Alice, sent by Intruder and received by Bob over channel \isacode{Recv}.

In the middle of the diagram, three functions are provided: 
\begin{enumerate*}[label=(\arabic*)]
    \item choose one event (1 to 3 in this case) from all enabled events to animate;
    \item reset manual animation so users can start from the beginning again; and
    \item choose the location of Eve for animation.
\end{enumerate*}
When users perform the animation, their interaction history will be dynamically displayed as a sequence diagram at the bottom. It is worth mentioning that the web interface supports multiple sessions, allowing users to animate the same protocol or different protocols simultaneously.


\Cref{fig:nswj3_web_auto} shows the automatic checking interface, where users can configure what property to check and its specific input. Now, three kinds of properties can be verified: secrecy, correspondence, and injective correspondence. For secrecy, users need to choose a channel for checking (such as Leak, Terminate, or Signal) and then optionally input the message in the input box. If no message is specified, the tool will check the reachability of all events on the chosen channel. For correspondence, users need to configure the event (event 1) for monitoring in addition to the event (event 2) for reachability checking. For example, the figure shows a configuration to check the authenticity of Bob in NSWJ3: check the reachability of the signal \isacode{EndProt} for Bob to finish the run with Alice using their nonces N0 and N1 while monitoring the occurrence of the signal \isacode{StartProt} for Alice to start the run with Bob using the same nonces. If the \isacode{EndProt} is reached without the occurrence of the signal \isacode{StartProt} before it, a counterexample is displayed. Then, users can click a button to view the counterexample in a sequence diagram. 

%
%
\Cref{fig:nswj3_secrecy_out_range_bob} shows a counterexample for the secrecy checking of NSWJ when Eve1 is chosen.
Alice's nonce (N0) is leaked because Eve is located outside Bob's jamming range. When Alice sends a watermarked message, Intruder can derive its plain message using the inference rule Wat1 in~\Cref{table:inference} and learn N0. Then Intruder leaks it into the environment. 
This is not surprising because the intruder can hear watermarked messages from Alice or Bob.
So, based on the watermarking rule in \Cref{table:inference}, the clear message is derived.
What is more surprising is authenticity, which holds even when secrecy is violated.
The original NSPK protocol is insecure, and both properties are violated~\cite{Lowe1996,Ye2024a}.
The clear explanation is that watermarking prevents authenticity from being compromised; in particular, though the intruder can learn the clear message, they cannot fake a watermarked message using Alice's or Bob's watermarking bitmask because they do not know these bitmasks.

Both our terminal and web interfaces allow users to do user-guided verification. For the web interface, users can manually explore a protocol for some steps and then want to know if there are any security issues after this stage. They can utilise the automatic verification function to conduct checks. In this case, the verification will start from the explored state instead of from the beginning. This could reduce the complexity of verifying large protocols and make them more scalable.

\ifdefined \CHANGES \indexprologue{%
  This index lists for each comment the pages where the text has been modified to address the comment. Since the same page may contain multiple changes, the page number contains the index of the change in superscript to identify different changes. Finally, the page number contains a hyperlink that takes the reader to the corresponding change.%
}%
\printindex[changes] \fi

\end{document}